\documentclass[11pt,a4paper]{article}
\usepackage{amsfonts}
\usepackage{amsmath}
\usepackage{amssymb}
\usepackage{bbold}
\usepackage{graphicx}
\usepackage{url}
\usepackage{bbm}
\usepackage{epsfig}
\usepackage[T1]{fontenc}
\usepackage[latin9]{inputenc}
\usepackage{multirow}
\usepackage{color}
\usepackage{caption}
\usepackage{subcaption}

\newcommand{\tb}{\bar t}

\newcommand{\qbp}{\bar q^{\prime}}

\newcommand{\ttbh}{ t \tb H}

\newcommand{\als}{\alpha_{\rm s}}
\newcommand{\shat}{\hat s}

\newcommand{\muf}{\mu_{\rm F}}
\newcommand{\mur}{\mu_{\rm R}}
\newcommand{\sigh}{\hat \sigma}

\newcommand{\nn}{\nonumber}

\def\beq{\begin{equation}}
\def\eeq{\end{equation}}
\def\bear{\begin{eqnarray}}
\def\eear{\end{eqnarray}}

\def\bet34{\beta_{kl}}

\begin{document}
\begin{center}
	{\Large \textbf{On soft gluon resummation for associated single top production with a Higgs boson at the LHC}}\\
	\vspace{.7cm}
	Anna Kulesza$^{a,}$\footnote{\texttt{anna.kulesza@uni-muenster.de}}, Laura Moreno Valero$^{a,}$\footnote{\texttt{lmore\_02@uni-muenster.de}},  and Vincent Theeuwes$^{b}$\footnote{\texttt{vtheeuwe@gmail.com}}

	\vspace{.3cm}
	\textit{
		$^a$ Institute for Theoretical Physics, WWU M\"unster, D-48149 M\"unster, Germany\\
		$^b$ Institute for Theoretical Physics, Georg-August-Univesity G\"ottingen, Friedrich-Hund-Platz 1, 37077 G\"ottingen, Germany\\
}
\end{center}   

\vspace*{2cm}
\begin{abstract}
We investigate threshold resummation of soft gluon corrections for the $s$-channel process of a single top quark production in association with a Higgs boson in the Standard Model. We choose to work in the three-particle invariant mass kinematics in direct QCD, i.e. in the space of Mellin moments. Our results take into account terms of up to next-to-leading logarithmic (NLL) precision, as well as  $\cal{O} (\als)$ non-logarithmic terms which do not vanish at threshold.  After presenting analytical expressions we discuss the corresponding numerical results for the total cross section and the invariant mass distributions at the LHC.
\end{abstract}


\clearpage
\setcounter{footnote}{0}


\section{Introduction}
\label{s:intro}

The Higgs boson interactions with top quarks and electroweak gauge bosons are one of the most important subjects of studies at the LHC. The strength of the top-Higgs Yukawa coupling can be probed directly without making any assumptions on its nature in the associated production processes involving both a top quark and a Higgs boson in the final state. Alongside the intensively studied associated production with a top-antitop quark pair, $t \bar t H$, the associated process of a single top quark and a Higgs boson ($t H$) production has been attracting a lot of attention lately. Although in the Standard Model (SM) the $tH$ production rate at the LHC is only one tenth of the rate for the $t \bar t H$ production ~\cite{Maltoni:2001hu},  the process is of great interest.  It offers a unique possibility to directly measure not only the value of the Yukawa coupling but also a relative sign between it and the coupling of a Higgs boson to a $W$ boson. The $tH$ cross section receives contributions from diagrams with $H$ radiated by the $t$ quark as well as from diagrams with $H$ radiated by the $W$ boson. The value of the cross section depends strongly on the interference between these diagrams. In fact, it is the interference between diagrams involving the two couplings that is responsible for the small value of the cross section in the SM.  For the same reason the $tH$ process offers a very sensitive probe of new physics interactions. The small cross section notwithstanding, first experimental searches for the process have been already conducted at the LHC~\cite{Khachatryan:2015ota,Sirunyan:2018lzm} and the rate for $H$ boson production in the $tH$ process is included in the measurements together with the $t \bar t H$ process~\cite{Aad:2020mkp, Aad:2020ivc, Sirunyan:2020icl}.

Given on one side the importance of the $tH$ production for new physics searches and, on the other side, much larger statistical samples and expected better understanding of the systematic uncertainties of the future LHC measurements, the precision with which the theoretical predictions are known becomes a central issue. At lowest order in perturbation theory, the production mechanisms for the associated production of the Higgs boson with a single top quark can be classified according to the virtuality of a $W$, which is either exchanged or emitted in the process. Correspondingly, the three mechanisms are the $t$-channel ($ qb \to t H q'$), $s$-channel ($ q \bar q' \to t H \bar b$) and the $tW$ associated channel ($g b \to t H W$). Strictly speaking, this classification is only valid assuming five active quark flavours, i.e. in the so-called five-flavour scheme (5FS) which involves $b$ quarks in the initial state. In the four-flavour scheme (4FS), initial state $b$-quarks can only appear as a result of a $g \to b \bar b$ splitting. First calculations of the next-to-leading order (NLO) QCD correction for the $t$-channel (in 4FS and 5FS) and $s$-channel production in the SM were reported in~\cite{Demartin:2015uha}, whereas an analysis of the NLO QCD corrections to the $t$-channel process in the framework of SM Effective Field Theory (SMEFT) was undertaken in~\cite{Degrande:2018fog}. Recently, NLO QCD+EW precision for the $tHj$ production was reached in the combined analysis of all channels in 5FS ~\cite{Pagani:2020mov}.  

With the full NNLO calculations being technically out of reach for the $ 2 \to 3$ processes considered here, a question remains if at least some part of the higher order corrections can be investigated already now. Higher-order corrections due to soft gluon emissions provide a class of corrections which can be identified and accounted for in a systematic manner, as shown in the case of the  $\ttbh$ or $t \bar t Z/W$ production~\cite{Kulesza:2015vda, Kulesza:2016vnq, Kulesza:2017ukk, Kulesza:2018tqz, Kulesza:2020nfh, Li:2014ula, Broggio:2015lya, Broggio:2016zgg, Broggio:2016lfj, Broggio:2017kzi, Broggio:2019ewu}. In particular, it has been reported in~\cite{Kulesza:2017ukk, Kulesza:2018tqz, Kulesza:2020nfh}, that adding soft gluon corrections calculated at the next-to-next-to-leading-logarithmic (NNLL) accuracy leads to a significant reduction of the scale uncertainty and a correction to the $\ttbh$ or $t \bar t Z$ total cross section of up to 20\%, depending on the choice of the central scale. Of course, one cannot expect a similar conclusion to automatically apply to the single associated top production, as for example the initial state and final state partons are different in both cases. In particular, in contrast to  the $tH$ production, the $\ttbh$ process involves the gluon-gluon channel, so it is reasonable to expect smaller impact of soft gluon corrections in the former case.   Nevertheless, since the processes are often analysed together by the experiments, theoretical predictions for the two processes should be known with the same accuracy. In a similar, in this respect, case of the associated heavy gauge boson production, the $t \tb W$ process, taking place at LO only in the quark channel, is known at the NLO+NNLL accuracy, i.e. the same level of accuracy as reached for the $t \tb Z$ process.

So far, only the predictions for the $t$-channel $tH$ production have been considered in the context of soft gluon emission.  More specifically,  in~\cite{Forslund:2020lnu,Forslund:2021evo} contributions to the NNLO corrections due to such emission were investigated. That study used the so-called one particle inclusive (1PI) kinematics. Performing full resummed calculations, i.e. obtaining numerical predictions which include soft gluon corrections to all order in perturbation theory in the direct resummation framework (i.e. in Mellin space) is known to be notoriously difficult in 1PI, see e.g~\cite{Hinderer:2018nkb}. For processes involving  jet production it has been achieved only for inclusive enough quantities~\cite{deFlorian:2007fv}. Notably, 1PI threshold resummation for single-inclusive jet production has been carried out in the SCET framework~\cite{Liu:2017pbb}.

In this work, we investigate the application of the direct Mellin resummation techniques to the $s$-channel $tH$ production. While the contribution from the $s$-channel to the $tH$ production is much smaller than the one from the $t$-channel, it proves a useful setting for an exploratory study. The $s$-channel does not involve initial state $b$-quarks and therefore it remains largely insensitive to the FS choice, which is not the case for the $t$-channel predictions. Given the known problems with performing fully resummed calculations in 1PI kinematics, we focus instead on the three-particle invariant mass kinematics, which has successfully been applied to the production of three massive particles, some carrying colour,  in the finals state~\cite{Kulesza:2015vda, Kulesza:2016vnq, Kulesza:2017ukk, Kulesza:2018tqz, Kulesza:2020nfh}. Application of these kinematics to the process in question requires considering kinematical information on a jet $j$ produced  together with a top quark and a Higgs boson.
This, in turn, brings up the question of an appropriate treatment of the jet. In the presence of particles other than massless partons in the final state at LO, 1PI kinematics allows to perfom calculations of observables which do not directly involve the jet by treating it inclusively, i.e. as unobserved. In other words, the threshold variable in 1PI kinematics does not directly involve the momenta of the massless parton. This is not the case in the invariant mass kinematics. Following~\cite{Kidonakis:1998bk}, in this kinematics the jet can be treated as massless or massive at the partonic threshold, leading to different structures of the logarithmic corrections. Massless jet approach results in double logarithmic contributions in the resummed factor describing the jet. Giving up the assumption of a vanishing jet mass at threshold leads to only single logarthmic contributions, albeit with coefficients depending on the jet size parameter. In the following, we consider the impact of threshold corrections on the prediction in both approaches. Our resummed results, take into account leading logarithmic (LL) and  next-to-leading logarithmic (NLL) terms together with ${\cal O} (\alpha_S)$ non-logarithmic contributuions which do not vanish at threshold.

The rest of the paper is structured as follows: in Section 2 we present the theoretical framework and the corresponding analytical expressions. Section 3 focuses on the numerical predictions. First we discuss the relation between the full NLO cross sections and the expansion of the resummed cross section truncated at NLO. Then we present the resummed NLL predictions matched to the NLO results. Section 4 contains a summary  of our results.

\section{NLL resummation in the triple invariant mass kinematics: theoretical framework}
\label{s:theory}

In the following we consider the corrections due to gluon emissons in the limit of the invariant mass $Q$ of the top quark ($t$), the Higgs boson ($H$) and the jet ($j$) becoming close the partonic center-of-mass energy $\shat$,  {$Q^2 \equiv  (p_t +p_H+ p_j)^2 \to \shat$. This corresponds to the limit $\hat \rho \to 1$ of the threshold variable $\hat \rho \equiv Q^2/\shat$. In the $s$-channel process, the final state jet is initated by the $b$-quark. In this work we treat the $b$ quark as massless, i.e. we use a jet function to describe the (quasi-)collinear emission from the final state quark. Further, we follow~\cite{Kidonakis:1998bk} regarding the treatment of the final state jet, and apart from an assumption of a massless jet with $p_j^2=0$, we also consider the case of a non-vanishing jet mass. For massless jets  $Q^2 =(p_t+p_H)^2 + 2 (p_t+p_H) \cdot p_j$. 

In general, the construction of the resummation formalism for a $2 \to 3$ process builds up on the earlier work for the $2 \to 2$ processes~\cite{Kidonakis:1998bk, Contopanagos:1996nh, Kidonakis:1998nf, Bonciani:2003nt}.  Here we adapt the expressions for the $t \bar t H (W,Z)$  from~\cite{Kulesza:2017ukk, Kulesza:2018tqz} to account for the presence of a final state jet instead of a top quark. The factorization of the soft emission, necessary to perform resummation, takes place in Mellin space, where the Mellin moments $N$ of the partonic cross sections are taken w.r.t. the variable $\hat \rho$.   At the next-to-leading logarithmic (NLL) accuracy, the resummed expression for the partonic cross section has a schematical form
\begin{eqnarray}
\tilde\sigh^{{\rm (NLL)}}_{i j}=  \mathrm{Tr} \left[\mathbf{H}_{ij \to tHk} \,\, \mathbf{S}_{ij \to tHk} \right] \Delta_i \Delta_{j} J_k \, 
\label{eq:res:fact_diag_NLL}
\end{eqnarray}

Since the soft radiation is coherently sensitive to the colour structure of the underlying hard process $i j \to t H k$,  the hard function $\mathbf{H}$, representing hard off-shell dynamics of the process, and the soft function $\mathbf{S}$, describing pure soft radiation, are matrices in color space. The $\Delta_i$ jet functions contain logarithmic contributions due to (soft-)colinear radiation from the incoming quarks or antiquarks, while $J_k$ corresponds to the final state jet function describing emissions of a quark $k$. 

In our calculations, following~\cite{Kidonakis:1998bk}, we define the final state jet four-momentum as the sum over the four-momenta of particles flowing into a cone of half-aperture (angular radius) $\delta$ around the jet axis. This approach corresponds to the small cone approximation~\cite{Furman:1981kf, Aversa:1989xw} used in jet calculations.  In the small cone approximation the $\delta$ parameter is related to the jet radius $R$ through $\delta=R/\cosh(\eta)$, where $\eta$ is the pseudorapidity of the jet. The approximation is known to work very well even up to relatively large values of $R \sim 0.7$~\cite{Jager:2004jh,Mukherjee:2012uz}.

The incoming jet functions are independent of the process and read
\begin{equation}
\Delta_i   =  \exp\left[\int_0^1 \text{d}z \, \frac{z^{N-1}-1}{1-z}  \left\{ \int^{(1-z)^2 Q^2}_{\mu_F^2}\frac{dq^2}{q^2} A_i (\als(q^2))\right\} \right]\,,
\end{equation}
where $\mu_F$ is the factorization scale. The outgoing jet function has a different form depending on how the jet is treated~\cite{Kidonakis:1998bk}. Specifically, if the jet is required to be massless it is given by 
\begin{equation}
J_k  =  \exp\left[\int_0^1 \text{d}z \, \frac{z^{N-1}-1}{1-z}  \left\{\int_{(1-z)^2 Q^2}^{(1-z) Q^2}\frac{dq^2}{q^2} A_k(\als(q^2))+\frac{1}{2} B_k(\als((1-z) Q^2))\right\}\right]\,.
\label{eq:jet:massless}
\end{equation}
Otherwise, for massive jets we have
\begin{equation}
J_k  = \exp\left[\int_0^1 \text{d}z \, \frac{z^{N-1}-1}{1-z}  C^{\rm M}_k \left(\als((1-z)^2 Q^2)\right)\right]\,.
\end{equation}
As discussed in~\cite{Kidonakis:1998bk}, the outgoing jet function contains double threshold logarithms if the jets are treated as massless, but only single logarithms in the case of a non-vanishing jet mass. Another important difference is that in the former case the jet function $J_k$ does not depend on the $\delta$ parameter, in contrast to the logarithmic dependence on $\delta$ which enters through the coefficient $C^{\rm M}_k$ in the latter case. The coefficients $A_i$, $B_i$, $C^{\rm M}_i$ are all perturbative series in $\als$,
$$
F_i = \frac{\als}{\pi} F_i^{(1)} + \frac{\als^2}{\pi^2} F_i^{(2)} + \dots\,,
$$
with $F_i$ standing in for any of the coefficients. In order to achieve the NLL accuracy, $A_i^{(1)}, A_i^{(2)}, B_i^{(1)}$ and $C_i^{\rm M, (1)}$ are needed. At LO, the $tH$ production involves only quarks in the initial and final state, i.e. $i=q$. The LL $A_q^{(1)}$ coefficient as well as the NLL $A_q^{(2)},\  B_q^{(1)}$ coefficients of the quark jet functions are well known~\cite{KT, CET}
\bear
A_q^{(1)} &=& C_F, \qquad A_q^{(2)} = {1 \over 2} C_F K\quad   {\rm with} \quad K=C_A\left(  {67 \over 18} -{\pi^2 \over 6}\right) -{5 \over 9} N_f \nn \\
B_q^{(1)} &=& \gamma_q = -\frac{3}{2} C_F.
\eear
For jets massive at threshold, our calculations return
\beq
C^{\rm M, (1)}_q = C_F \log \left( {Q^2 \over \delta^2 E_q^2} \right) \,,
\eeq
with the $\delta$ dependence entering logarithmically, as expected.

The function $\mathbf{S}$ sums logarithmic contributions from soft wide-angle emission. In Eq.(\ref{eq:res:fact_diag_NLL}) it is already rescaled by the inverse of the soft function contributions for the incoming and outgoing lines, to avoid double counting of soft contributions included in the $\Delta_i$ and $J_k$ functions used in this work. In the most general case, the solution of the evolution equation  for $\mathbf{S}$~\cite{Kidonakis:1998bk} involves path-order exponentials of the integrals over the soft anomalous dimension $\mathbf{ \Gamma} ={ \als \over \pi} \mathbf{\Gamma}^{(1)} + \dots$. To reach the NLL accuracy, we need to know the one-loop soft anomalous dimension $\mathbf{ \Gamma}^{(1)}$.  For the process $q \qbp \to t H k$ with $q, \qbp,k$ corresponding to massless quark lines we find  
\begin{eqnarray}
\Gamma^{(1)}_{q \qbp \rightarrow t H k}&=&C_F\mathbbm{1} \left[\log\left(\frac{s_{t k}-m_t^2}{m_t Q}\right)-\frac12\left(1-i\pi\right)\right]\nonumber\\
&&-\frac12 \boldsymbol{C_I}\left[\log\left(\frac{s_{tk}-m_t^2}{m_t Q}\right)-\log\left(\frac{t_{qk}\left(t_{\qbp t}-m_t^2\right)}{m_t Q^3}\right)\right]\nonumber\\
&&-\mathbf{T}_{qt}\log\left(\frac{t_{\qbp k}\left(t_{q t}-m_t^2\right)}{t_{q k}\left(t_{\qbp t}-m_t^2\right)}\right) \,,
\label{eq:SAD}
\end{eqnarray}
where we have $s_{ij}=(p_i+p_j)^2$,  $t_{ij}=(p_i-p_j)^2$, $\boldsymbol{C_I}=\text{Diag}\left(0, C_A\right)$ and
\begin{eqnarray}
\mathbf{T}_{qt}&=&
\begin{pmatrix}
0 & -\frac{C_F}{2C_A} \\
-1 & -2C_F+\frac{C_A}{2} 
\end{pmatrix}
\end{eqnarray}
or
\begin{eqnarray}
\mathbf{T}_{q\bar{t}}&=&
\begin{pmatrix}
0 & \frac{C_F}{2C_A} \\
1 & 2C_F-C_A 
\end{pmatrix}\,
\end{eqnarray}
for the $q \qbp \to \tb H k$ process. Our results  for  $\mathbf{ \Gamma}^{(1)}$  are obtained in the singlet-octet $s$-channel basis and agree with the result recently published in the literature~\cite{Forslund:2020lnu}. 
 
Apart from knowing the soft anomalous dimension, evaluation of the soft function at any scale also requires  knowing the boundary condition $\mathbf{\tilde S}$   in the solution of the renormalization group equation at $\mu_R = Q/\bar N$, with $\bar N = N e^{\gamma_E}$ and $\gamma_E$ the Euler constant, cf. Eq.~(\ref{eq:soft:R:evol}) below.  $\mathbf{\tilde S}$ is a purely eikonal function which can be calculated perturbatively
\beq
\mathbf{\tilde S}= \mathbf{\tilde S}^{(0)} + \frac{\als}{\pi} \mathbf{\tilde S}^{(1)} + \dots
\label{eq:softexp}
\eeq
The leading-order soft function is given by 
\begin{eqnarray}
\mathbf{\tilde S}^{(0)}=
\begin{pmatrix}
C_A^2 & 0 \\
0 & \frac{C_A C_F}{2} \,
\end{pmatrix}.
\end{eqnarray}
With $\mathbf{\tilde S}^{(0)}$ and $\mathbf{\Gamma}^{(1)}$, NLL precision can be reached. Any higher precision  requires also the expression for  $\mathbf{\tilde S}^{(1)}$. For massless initial state \mbox{$i=q,\qbp$}  and final state $k$ quarks we obtain finite contributions in the form
\begin{eqnarray}
\mathbf{ \tilde S}^{(1)}&= &\mathbf{\tilde S}^{(0)} \left\{ C_F\mathbbm{1}\left[S_{tt}-2S_{tk}\right] +\boldsymbol{C_I}\left[S_{q\qbp}+S_{tk}-S_{qk}-S_{\qbp t}\right] \right. \nonumber\\
&&+\left. 2\mathbf{T}_{qt}\left[S_{qt}+S_{\qbp k}-S_{qk}-S_{\qbp t}\right] \right\}
\end{eqnarray}
with
\begin{eqnarray}
S_{q \qbp}&=&-\frac{\pi^2}{12}\nonumber\\
S_{tt}&=&\frac{1}{2\beta_t}\log\left(\frac{1+\beta_t}{1-\beta_t}\right)\nonumber\\
S_{tk}&=&- \frac{\pi^2}{24}+ \frac{1}{8} \log^2 \frac{1+\beta_t}{1-\beta_t} + \frac{1}{2} \left\{ \text{Li}_2 \left(1- \frac{2E_t E_k (1 -\beta_t)}{s_{tk}-m_t^2}\right) \right. \nn \\
&&\left. +\text{Li}_2 \left(1- \frac{2E_t E_k (1 +\beta_t)}{s_{tk}-m_t^2}\right) \right\}\nonumber\\
S_{it}&=&- \frac{\pi^2}{24}+ \frac{1}{8} \log^2 \frac{1+\beta_t}{1-\beta_t}  + \frac{1}{2} \left\{ \text{Li}_2 \left( 1 - \frac{E_t Q(1 +\beta_t)}{ m_t^2 - t_{it}}\right)\ \right. \nn \\
&& \left. +\text{Li}_2 \left( 1 - \frac{E_t Q(1 -\beta_t)}{ m_t^2 - t_{it}}\right) \right\}\nonumber\\
S_{ik}&=&- \frac{\pi^2}{12}  + \frac{1}{2}\text{Li}_2 \left( 1 + \frac{2E_k Q}{t_{ik}}\right)
\end{eqnarray}
and  $\beta_t=\sqrt{1-m_t^2/E_t^2}$.

Calculations of the soft function in the dimensional regularization (DR) scheme leading to the above results deliver also information on double and single poles in $\epsilon$. The one-loop soft function corresponding to $\mathbf{\tilde S}^{(1)}$ does not contain $1/\epsilon^2$ poles, while the coefficients of the $1/\epsilon$ terms return the soft anomalous dimension given in Eq.~(\ref{eq:SAD}), thus providing an additional check of our calculations.

In order to simplify the solution of the renormalization group equation for the soft function, and, correspondingly, the expression in eq.~(\ref{eq:res:fact_diag_NLL}), we write it in the color basis in which the one-loop soft anomalous dimension $\mathbf{ \Gamma}^{(1)}$, driving the evolution of the soft-function, is  diagonal. We denote all matrices in this basis with the subscript $R$.  The $IJ$ element of the diagonalised one-loop soft anomalous dimension is thus
$$
\mathbf{\Gamma}^{(1)}_{R, IJ} = \lambda_I^{(1)} \delta_{IJ}
$$
where $\lambda_I^{(1)}$ are the eigenvalues of $\mathbf{\Gamma}^{(1)}$. The soft function has then the form 
\beq
\mathbf{S}_{R, IJ} = \mathbf{\tilde S}_{R, IJ} \exp\left[\frac{\log(1-2\lambda)}{2 \pi b_0}
\left(\left( \lambda^{(1)} _{I}\right)^{*}+\lambda^{(1)} _{J}\right)\right]\, ,
\label{eq:soft:R:evol}
\eeq
with $\lambda= \als(\mu_R^2) b_0 \log N$, $b_0= (11C_A-2 nf)/(12\pi)$ and $\mu_R$ the renormalization scale.

The NLL precision of the resummed expression can be increased by including the ${\cal O}(\als)$ terms in the expansions of the hard function
$\mathbf{H} = \mathbf{H}^{\mathrm{(0)}} + \frac{\als}{\pi}  \mathbf{H}^{\mathrm{(1)}} +\dots$
and the soft function $\mathbf{\tilde S}$, cf. Eq.~(\ref{eq:softexp}), as well as collinear non-logarithmic contributions. In the  literature on direct QCD resummation such terms are often collectively referred to as the "${\mathbf C^{(1)}}$" coefficient. 
We will therefore refer to the precision obtained by adding the information on ${\mathbf C^{(1)}}$ to the NLL result as "NLLwC"~\footnote{This accuracy is formally equivalent to the "NLL'" notation used in the SCET literature on resummation.} and give the corresponding expressions below. In the $R-$basis we have
\begin{eqnarray}
 \frac{d \tilde\sigh^{{\rm (NLLwC)}}_{ij }}{dQ^2}&&\hspace{-0.9cm}(N,Q^2,\{m^2\},\muf^2, \mur^2) =  \label{eq:res:fact_diag_NLLwC_R} \\ 
 &&\hspace{-0.9cm} \ (\mathbf{H}_{R,IJ} \mathbf{\tilde S}_{R,JI})(Q^2, \{m^2\},\muf^2, \mur^2, \delta) \nn \\ 
&\times&
\Delta_i(N+1, Q^2,\muf^2, \mur^2) \Delta_j(N+1, Q^2,\muf^2, \mur^2) \nonumber \\
&\times& J_k(N+1, \delta, Q^2, \mur^2 )\, \exp\left[\frac{\log(1-2\lambda)}{2 \pi b_0}
\left(\left( \lambda^{(1)} _{J}\right)^{*}+\lambda^{(1)} _{I}\right)\right], \nonumber 
\end{eqnarray}
where $\{m^2\}$ stands for a set of mass parameters characteristic for the process.
The NLLwC precision is achieved with
\begin{eqnarray}
\mathbf{H}_{R}\, \mathbf{\tilde S}_{R}& = &\mathbf{H}^{\mathrm{(0)}}_{R} \mathbf{\tilde S}^{\mathrm{(0)}}_{R} \left( \mathbbm{1} + \frac{\als}{\pi} \ \mathbf{C}^{(1)}_{R} \right) \label{eq:hardXsoft} \\
&=& \mathbf{H}^{\mathrm{(0)}}_{R} \mathbf{\tilde S}^{\mathrm{(0)}}_{R} + \frac{\als}{\pi}\left[ \mathbf{H}^{\mathrm{(0)}}_{R} \mathbf{\tilde S}^{\mathrm{(1)}}_{R} + \mathbf{V}^{\mathrm{(1)}}_{R} \mathbf{\tilde S}^{\mathrm{(0)}}_{R}+ \mathbf{H}^{\mathrm{(0)}}_{R} \mathbf{\tilde S}^{\mathrm{(0)}}_{R} \left( 2 \tilde J_{in}^{(1)} + \tilde J_{out}^{(1)}  \right) \right]  \nn
\end{eqnarray}
If only the  $\mathbf{H}^{\mathrm{(0)}}_{R} \mathbf{\tilde S}^{\mathrm{(0)}}_{R}$ term, with its trace corresponding to the  leading order partonic cross section, is kept in the expression above, the precision of the resummed cross section in Eq.~(\ref{eq:res:fact_diag_NLLwC_R}) reduces to NLL.
Eq.~(\ref{eq:hardXsoft}) defines the NLLwC precision as the NLL precision supplemented with additional information included in the coefficient $ \mathbf{C}^{(1)}$. It consists of ${\cal O(\als)}$ contributions from the virtual corrections  $\mathbf{V}^{\mathrm{(1)}}$, collinear non-logarithmic contributions ${\tilde J}_{in}^{(1)}$ from incoming quark jets, as well as ${\cal O(\als)}$ non-logarithmic contributions ${\tilde J}_{out}^{(1)}$ from the outgoing jets. The virtual corrections $\mathbf{V}^{\mathrm{(1)}}$ are calculated numerically using the  aMC@NLO code~\cite{Alwall:2014hca}, whereas the collinear contributions are calculated analytically. In particular, for the outgoing quark jet with zero invariant mass we find the finite non-logarithmic contributions of the form
\begin{eqnarray}
{\tilde J}_{out}^{(1)} &=&C_F \left\{\frac{1}{4}\log^2 \frac{\mu_R^2}{Q^2}+\frac{3}{4}\log \frac{\mu_R^2}{Q^2}+ \frac{7}{4} + \frac{3}{4}\gamma_E - \frac{1}{2} \gamma_E^2 - \frac{ 5\pi^2}{12}
  - \log^2 \frac{2}{\delta} \right. \nn \\
& -& \log^2 \frac{E_k \, \delta}{Q} 
+    \frac{1}{4} \Bigg[ \frac{Q^2}{Q^2 + E_k^2 \bar{N} \delta^2} + 3 \log \left(1+ \frac{Q^2}{E_k^2 \bar{N} \delta^2}\right) \nonumber \\
&+ &2 \log^2 \bigg(1+ \frac{Q^2}{E_k^2 \bar{N} \delta^2}\bigg) + 4 \text{Li}_2 \left( \frac{Q^2}{Q^2+E_k^2 \bar{N} \delta^2} \right) \nonumber \\ 
&+ &  \left. \left. \frac{\pi^2}{3}\frac{2\,Q^6 + 5 \,  E_k^2 \, \bar{N} \, Q^4 \, \delta^2 + 4 \, E_k^4 \, \bar{N}^2 \, Q^2\, \delta^4}{2\, (Q^2+E_k^2 \bar{N} \delta^2)^3} \right] \right\} \,,
\label{eq:jet1massless}
\end{eqnarray}
whereas in the massive case we have
\begin{eqnarray}
{\tilde J}_{out}^{{\rm M},(1)} =  C_F  \hspace{-1.5em} && \left(  2 -  \frac{\pi^2}{6} + \frac{1}{4}\log^2 \frac{\mu_R^2 }{Q^2} + \frac{3}{2}\log \frac{\mu_R}{\delta E_k} -\log^2 \frac{2}{\delta}  + \log^2 \frac{E_k \delta} {Q} \right. \nn \\
&&- \left.  2\gamma_E \log \frac{Q}{\delta E_k} \right)
\label{eq:jet1massive}
\end{eqnarray}
where $k$ corresponds to a massless final state quark.
In Eq.~(\ref{eq:jet1massless}), apart from terms of ${\cal O} (1)$ in $N$, we show additional contributions formally subleading in $N$ that, as we discuss in the next section, turn out to be relevant numerically. These contributions given in the square bracket arise for $\bar N \sim Q^2/(E_k^2 \delta^2)$ and  vanish in the limit  $\bar N \gg Q^2/ (E_k^2 \delta^2)$.

Our calculations of jet functions, leading to Eqs.~(\ref{eq:jet1massless}) and~(\ref{eq:jet1massive}), return coefficients of  the $1/\epsilon^2$ and $1/\epsilon$ poles corresponding to cusp and non-cusp contributions to the jet  anomalous dimension. They also reproduce the same coefficients of the double and single logarithms of $N$ as given by the first-order expansion of a massless jet function, cf. Eq.~(\ref{eq:jet:massless}). For a massive jet function they return only a single logarithm of $N$, accompanied by a coefficient logarithmic in $\delta$, in agreement with literature~\cite{Kidonakis:1998bk}.

The numerical predictions in the next section for the resummation-improved hadronic cross sections for the $s$-channel process $h_1 h_2 \to t H +X $  are obtained through matching the NLO cross section with the hadronic equivalent of Eq.~(\ref{eq:res:fact_diag_NLLwC_R}). In this way, the full information from the NLO calculations is preserved, while  double counting of the LO and NLO contributions included in the NLLwC expression avoided. More specifically,
\bear
\label{hires}
\frac{d\sigma^{\rm (NLO+NLLwC)}_{h_1 h_2 }}{d Q^2}(Q^2,\{m^2\},\muf^2, \mur^2) &=& 
\frac{d\sigma^{\rm (NLO)}_{h_1 h_2 }}{d Q^2}({Q^2},\{m^2\},\muf^2, \mur^2) \\ \nn
&+&   \frac{d \sigma^{\rm
		(res-exp)}_
	{h_1 h_2 }}{d{Q^2}}( Q^2,\{m^2\},\muf^2, \mur^2) 
\eear
with
\bear
\label{invmel}
&& \!\!\!\!\!\! \frac{d \sigma^{\rm
		(res-exp)}_{h_1 h_2 }}{d{Q^2}} ({Q^2},\{m^2\},\muf^2, \mur^2) \! = \!\!\! \sum_{i,j=\{q, \qbp\}}\,
\int_{\sf C}\,\frac{dN}{2\pi
	i} \; \rho^{-N} f^{(N+1)} _{i/h{_1}} (\muf^2) \, f^{(N+1)} _{j/h_{2}} (\muf^2) \nn \\ 
&& \!\!\!\!\!\! \! \times\! \left[ 
\frac{d \tilde\sigh^{\rm (NLLwC)}_{q \qbp}}{d Q^2} (N,Q^2,\{m^2\},\muf^2, \mur^2) 
-  \frac{d \tilde\sigh^{\rm (NLLwC)}_{q \qbp}}{d Q^2} (N, Q^2,\{m^2\},\muf^2, \mur^2)
{ \left. \right|}_{\scriptscriptstyle({\rm NLO})}\, \! \right], \nn \\
\eear
where $ d \tilde\sigh^{\rm (NLLwC)}_{q \qbp} / dQ^2 |_{\scriptscriptstyle({\rm NLO})}$ represents the perturbative expansion of $ d \tilde\sigh^{\rm (NLLwC)}_{q \qbp} / dQ^2$ truncated at the same order of $(\als)$ as in the NLO result.
The Mellin moments of the parton distribution functions are defined in  the usual way as 
$$
f^{(N)}_{i/h} (\muf^2) = \int_0^1 dx \, x^{N-1} f_{i/h}(x, \muf^2)\,,
$$
and the inverse Mellin transform is evaluated using a contour ${\sf C}$ in the complex $N$ space using the minimal prescription method developed in~\cite{Catani:1996yz}. The total cross sections are calculated by integrating the invariant mass distribution in Eq.~(\ref{hires}) over $Q^2$.

\section{Numerical results }
\label{s:results}
In this section we discuss the results for NLO+NLL resummation of soft gluon  correction for the $s$-channel $tH$ production at the LHC collision energy $\sqrt S$=13 TeV. The NLO cross section is calculated using the aMC@NLO code~\cite{Alwall:2014hca}. We use  $m_t = 173$ GeV and $m_H = 125$ GeV and employ PDF4LHC15 parton distribution sets~\cite{Butterworth:2015oua, Dulat:2015mca, Harland-Lang:2014zoa, Ball:2014uwa, Gao:2013bia, Carrazza:2015aoa}, apart from LO results which are calculated with the LO MMHT2014 set~\cite{Harland-Lang:2014zoa}. Unless otherwise stated, the results shown in the plots are obtained for the central scale choice $\mu_{R,0}=\mu_{F,0}=\mu_0=Q$ and for the jet radius $R=0.6$. 

We begin by investigating how well an expansion of the resummed cross section approximates the NLO cross section. More precisely, we compare the expansion with the contribution to the NLO cross section originating from the $q \bar q'$ channel, i.e. the NLO cross section with the $qg$ channel removed, which we call "NLO (no qg)". Since the $qg$ channel is subleading in powers of $N$ w.r.t. the $ q \bar q'$ channel, it cannot be resummed within the current formalism and therefore should not be taken into consideration in the comparison. Given that the threshold definition we use here depends on the variable $Q$, in the first step we study differential distributions in $Q$ for a phenomenologically viable value of the $R$ parameter, $R=0.6$. In Fig.~\ref{fig1r06} we compare the NLO (no qg) distribution with the expansion of the NLL cross section up to NLO, called "NLLwC$|_{\rm NLO}$". 
Additionally, in the massless case we also show distributions calculated without power suppressed terms in $N$, cf. Eq.~(\ref{eq:jet1massless}). 
It is immediately clear from Fig.~\ref{fig1r06}(b) that the power suppressed corrections play a vital role. Therefore  from now on in all results where the final state jet  is treated as massless we will include these contributions. For most of the range of $Q$ shown in the plots, the expansion overestimates the NLO (no qg) distribution by around a few percent. Only at the lower end of the spectrum, the expansion underestimates the NLO result but this region is relatively small and does not provide a dominant contribution to the total cross section. We observe that the expanded results are quantitatively very similar independent of the assumption on the jet mass. It turns out that this is also a manifestation of the numerical importance of the power suppressed contributions. Namely, we find that the analytical expressions for  NLLwC$|_{\rm NLO}$ cross sections agree between the two approaches  in the limit $\delta^2 \ll 1/\bar N$, provided the power suppressed  contributions are included. Correspondingly, this behaviour also implies a relatively low relevance of large-$N$ terms for the considered process.

The lower panels of the plots in Fig.~\ref{fig1r06} display the ratio of the invariant mass distributions to the LO distribution. Firstly we notice that the NLO (no qg) corrections are substantial for all values of $Q$ and can be well above 50\% for the $Q$ values corresponding to the peak of the distribution. In comparison with the ratios for the NLO (no qg) distributions, the ratios of the NLLwC$|_{\rm NLO}$  to the LO distributions are flatter. This corresponds to the NLLwC$|_{\rm NLO}$ distributions closer in shape to the LO distributions than the NLO (no qg) distributions, in accordance with the former ones being calculated in the 3-particle kinematics limit.

\begin{figure}[t!]
\centering
\includegraphics[width=0.49\textwidth]{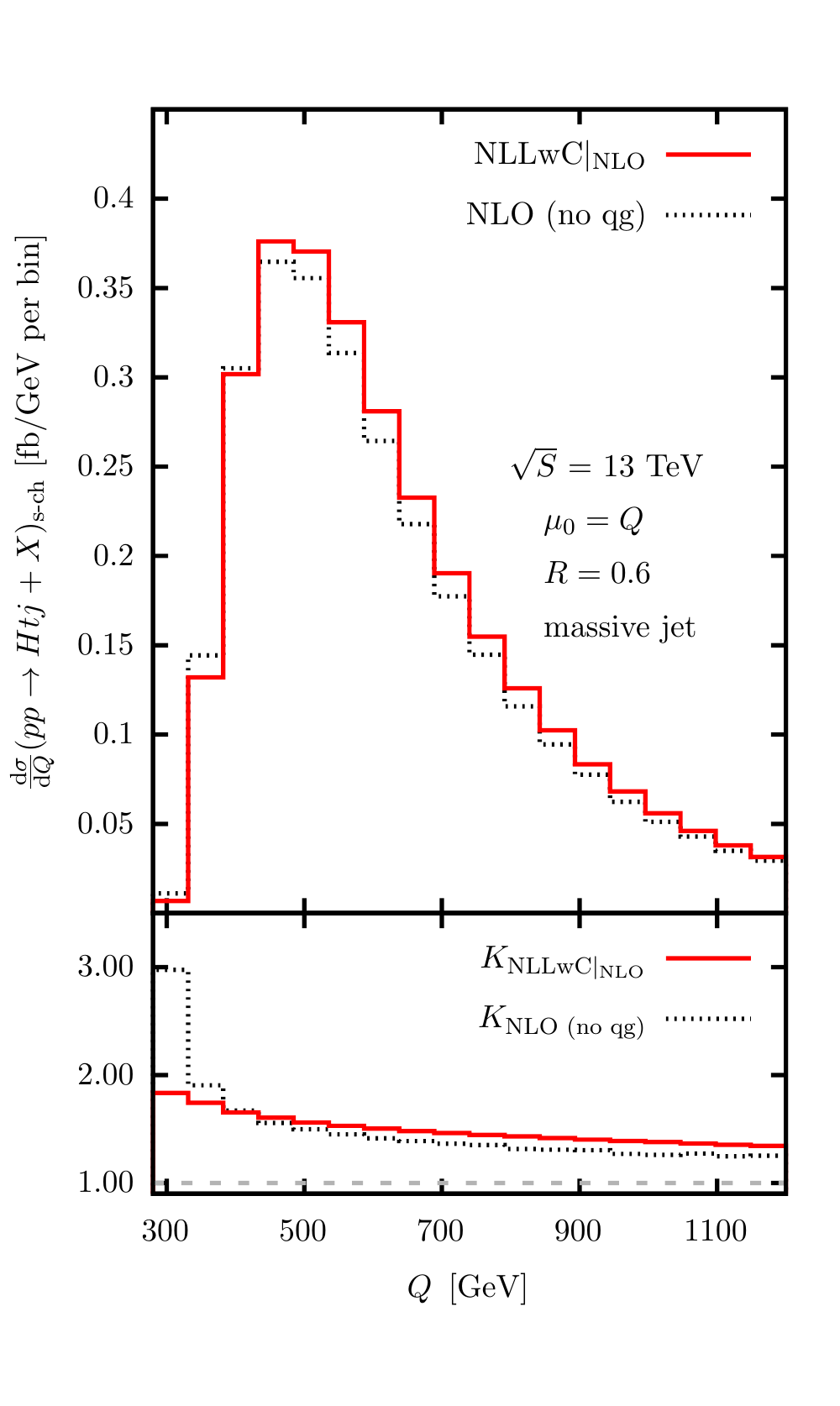}
\includegraphics[width=0.49\textwidth]{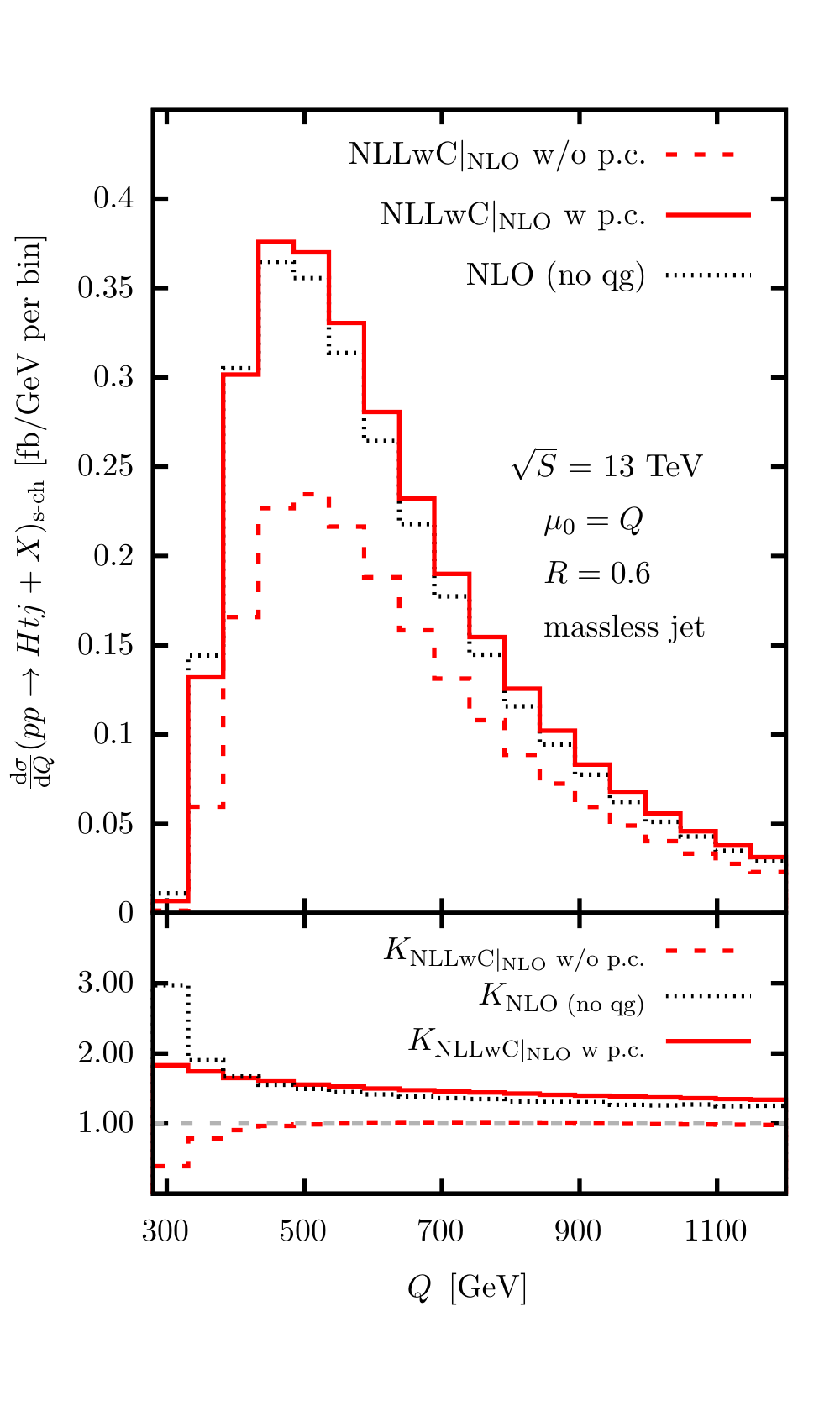}
\caption{Differential distributions in $Q$ for massive (left plot) and massless (right plot) jets  for $R=0.6$: (top) absolute values of the NLO(no qg), NLLwC|NLO distributions and (bottom) ratios of NLLwC|NLO and NLO(no qg) results to the LO predictions.}
\label{fig1r06}
\end{figure}

In Fig.~\ref{fig1r01} we study the invariant mass distributions obtained for a smaller value of $R$, $R=0.1$. For smaller $R$, the NLO (no qg) distribution in $Q$ naturally becomes narrower since narrower jets are less massive and the invariant mass of the full system decreases correspondingly. This can be also observed by comparing lower panels of Figs.~\ref{fig1r06} and~\ref{fig1r01}. Additionally, we see that the quality of the approximation of the NLO (no qg) result decreases slightly in comparison to the $R=0.6$ case. The dependence of the NLLwC$|_{\rm NLO}$ results on $\delta$ manifests itself in the higher absolute values for these distributions at smaller $R$. This dependence can be seen as an artifact of our setup, where by choosing the threshold variable dependent on the invariant mass $Q$, we are forced to define a jet, thus introducing the dependence on $\delta$ in the  NLLwC$|_{\rm NLO}$ cross sections. The dependence must cancel against $\delta$ dependence of the contributions from hard radiation, not included in the soft approximation, in the inclusive NLO cross section. At this very small value of $R$ some of the  the $\delta$-dependent terms can be very big individually, as can be seen in the right plot in Fig.~\ref{fig1r01} by comparing results with and without power subleading terms in $N$ in the massless jet case.

\begin{figure}[t!]
\centering
\includegraphics[width=0.49\textwidth]{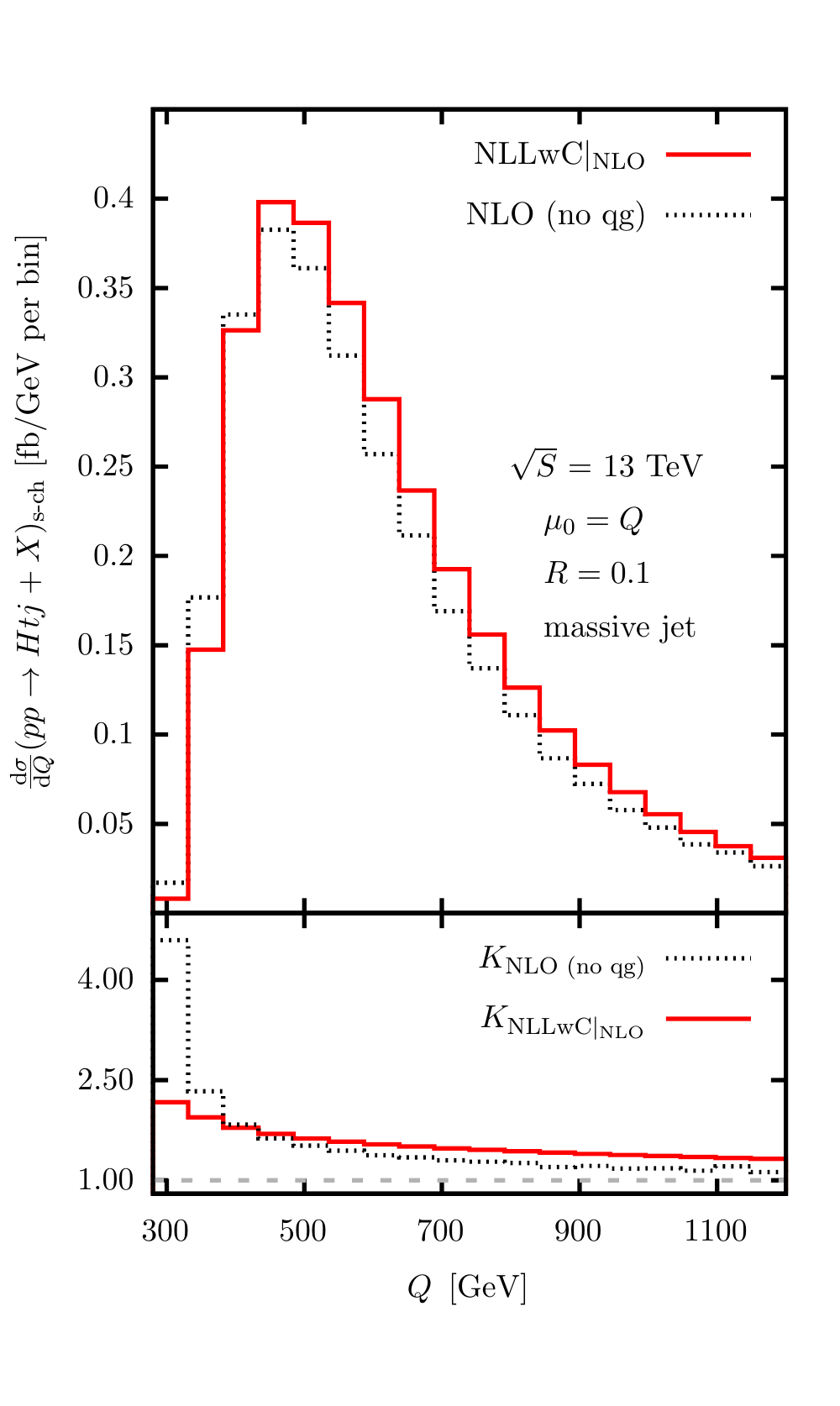}
\includegraphics[width=0.49\textwidth]{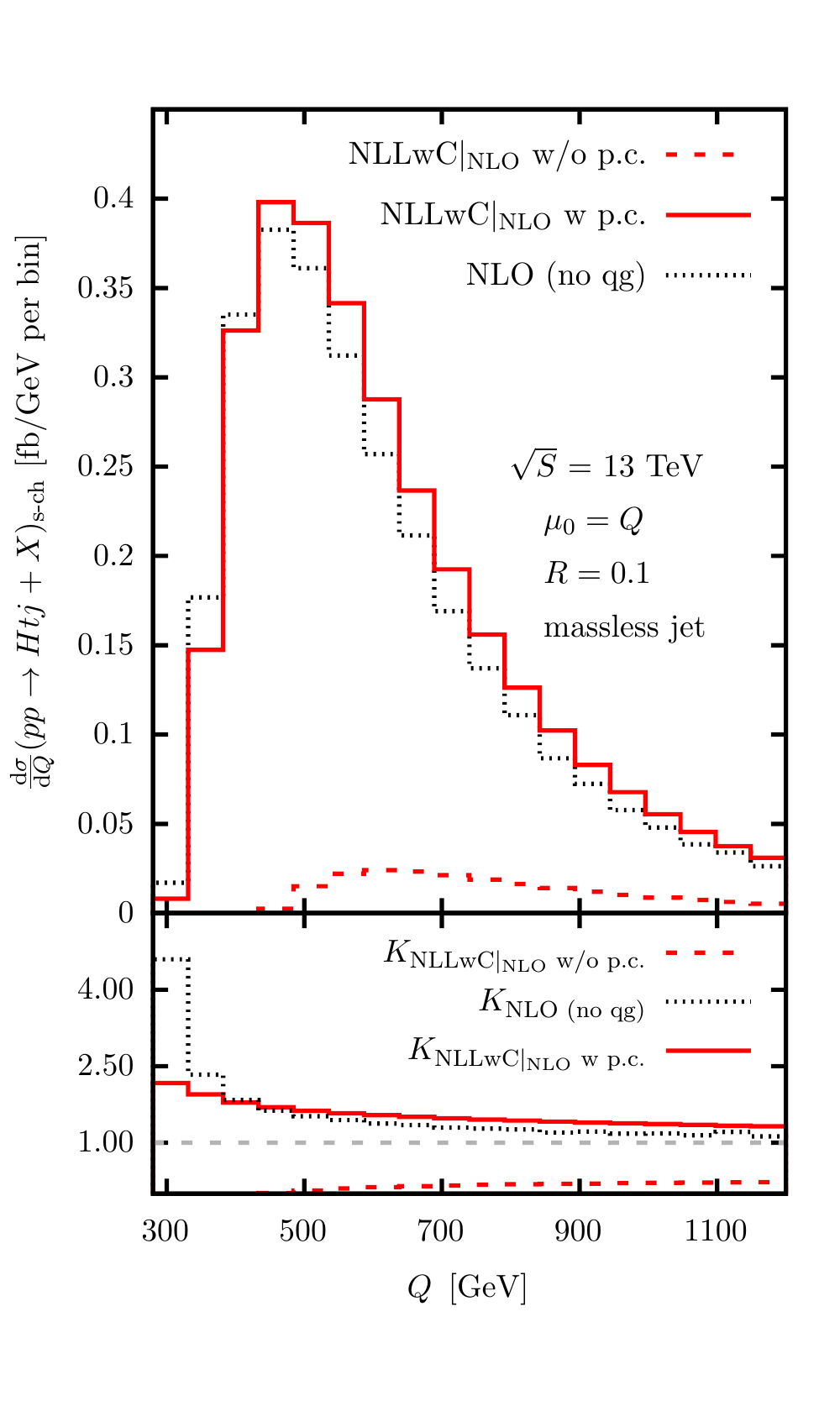}
\caption{Differential distributions in $Q$ for massive (left plot) and massless (right plot) jets for $R=0.1$: (top) absolute values  of the NLO(no qg), NLLwC|NLO distributions and (bottom) ratios of NLLwC|NLO and NLO(no qg) results to the LO predictions).}
\label{fig1r01}
\end{figure}

In Fig.~\ref{fig2} we show scale variation for $\mu=\mu_F=\mu_R$ around the central scale $\mu_0=Q$ of a set of $K$ factors, defined as the ratios of the NLO, NLO (no qg) and NLLwC$|_{\rm NLO}$ total cross sections to the LO cross section. First we observe that the full NLO corrections to the total cross section are substantial, above 35\% within the range of $\mu$ considered here. Fig.~\ref{fig2} also shows that most of the NLO cross section is provided by the quark channel, while the NLO $qg$ terms at $\mu/\mu_0=1$ provide contributions of around 5\% to the full NLO result or of around 20\% to the NLO corrections. These contributions are negative for the most of the scale range shown, specifically  where the difference between NLO (no qg) and NLO is the most pronounced. As already discussed, in our approach resummation is performed for the dominant $q\qbp$ channel, while the $qg$ channel is included at the NLO accuracy.

The overestimation of the NLO (no qg) results by the  NLLwC$|_{\rm NLO}$  approximation for most values of $Q$ observed for differential distributions carries over to the total cross sections. Consistently with differential results, we also observe that the numerical values of the NLLwC$|_{\rm NLO}$ expansions agree very well between results for massless and massive jets. The expansions overestimate the NLO (no qg) result by around 3\% at $\mu=\mu_0=Q$. In terms of $K$-factors, this corresponds to $K_{\rm NLO\;(no\;qg)}$ of around $1.45$ at this value of $\mu$, while the expansions provide $K$-factors of around 1.5, see Fig.~\ref{fig2}. The agreement among the $K$-factors improves significantly as the scale increases, indicating that the scale-dependent terms provide considerable contributions to the NLO (no qg) cross section. 

\begin{figure}
\centering
\includegraphics[width=0.55\textwidth]{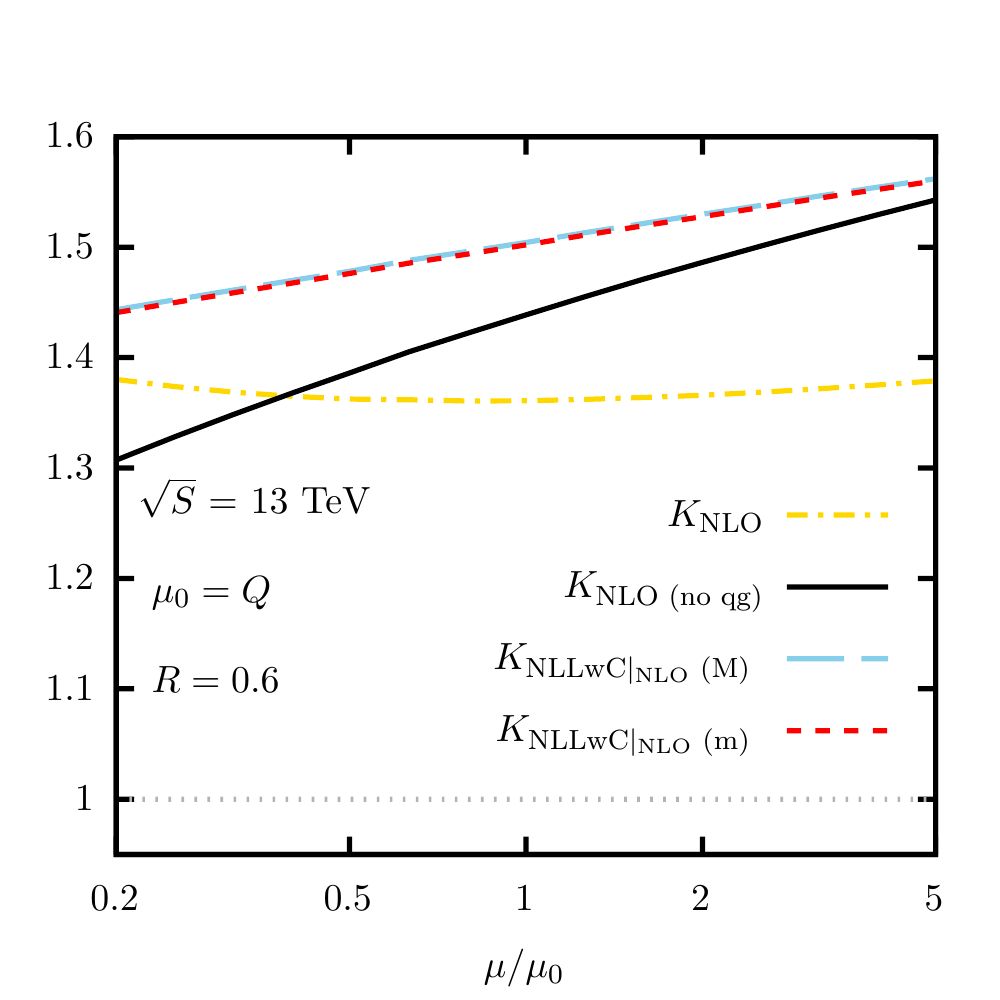}
\caption{Total $K$-factors as a function of $\mu/\mu_0$ for $R=0.6$: ratios of the total cross sections calculated at the NLO, NLO(no qg) and NLLwC|NLO accuracy to the LO total cross section with the final state jet treated as massive (M) or massless (m).}
\label{fig2}
\end{figure}

The dependence of the NLO expansions of the resummed cross section on the jet parameter $R$ is plotted in Fig.~\ref{fig3} for the same $K$-factors as shown in Fig.~\ref{fig2}. As $R$ gets smaller, the terms logarithmic in $\delta$ become more and more relevant, driving the difference between the expansion of the resummed cross section and the NLO result further apart. However, the dependence on $R$ is relatively mild for moderate $R$ values. Overall, for the central scale choices $\mu_0=Q$ and a wide range of $R$ values, the NLO(no qg) result differs from the NLLwC$|_{\rm NLO}$ expansion by up to a few percent.

\begin{figure}[t!]
\centering
\includegraphics[width=0.55\textwidth]{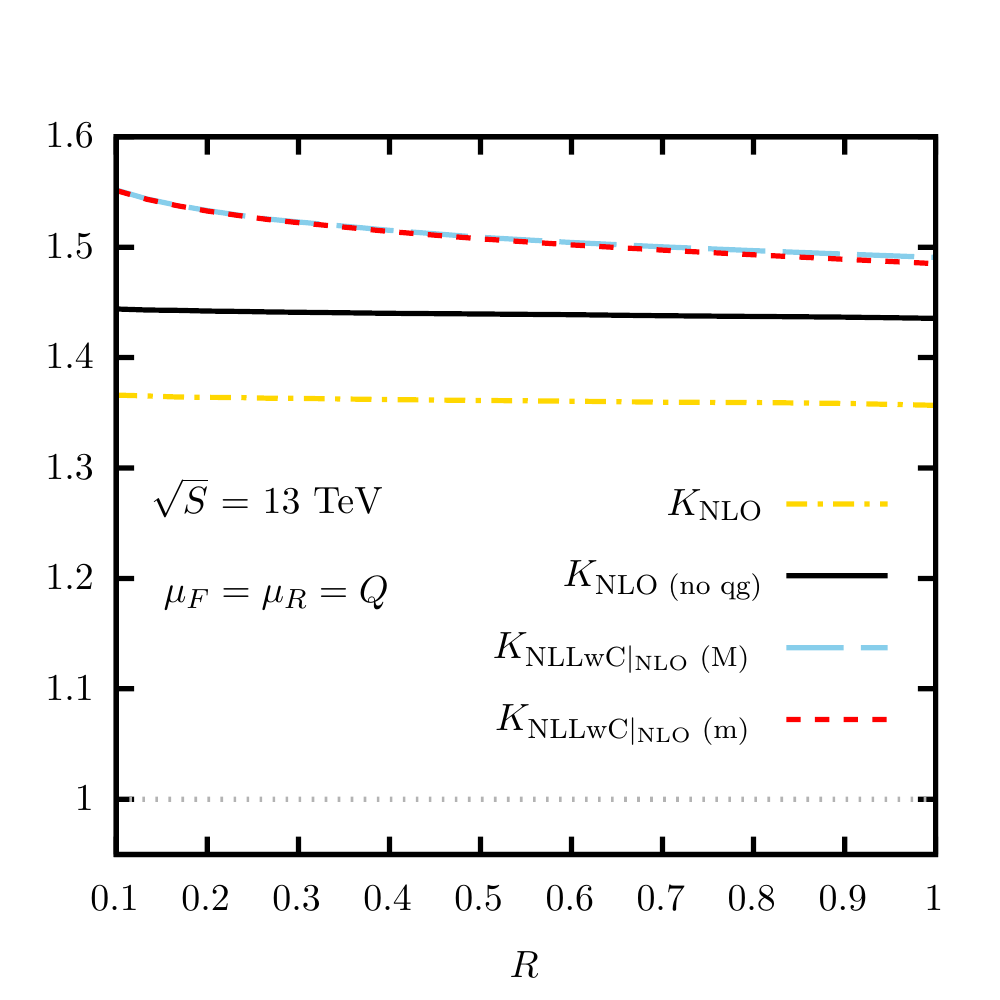}
\caption{Total $K$-factors as a function of $R$ at $\mu=\mu_0=Q$: ratios of the total cross sections calculated at the NLO, NLO(no qg) and NLLwC|NLO accuracy with massive (M) and massless (m) jets to the LO total cross section.}
\label{fig3}
\end{figure}

\begin{figure}[h!]
\centering
\includegraphics[width=0.49\textwidth]{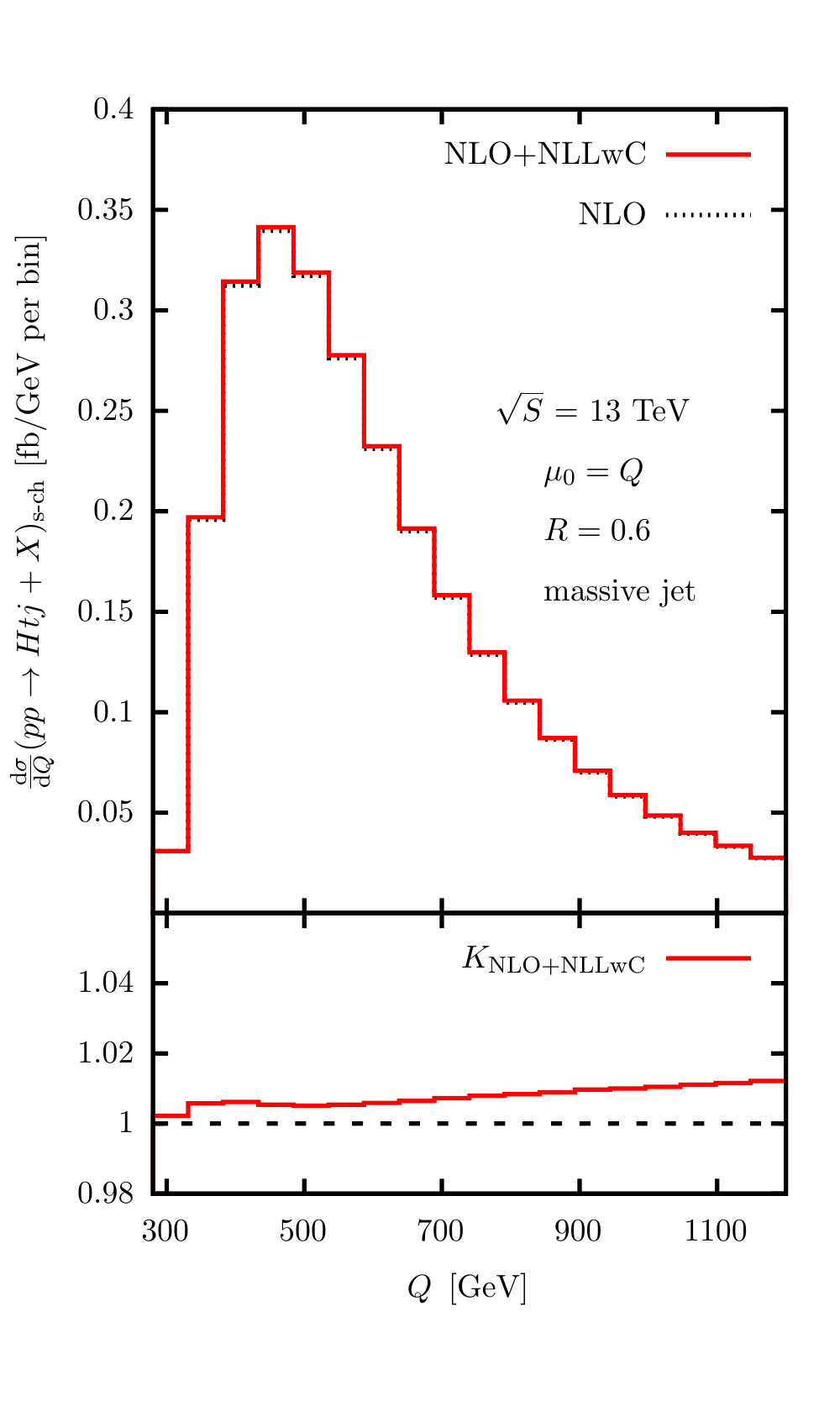}
\includegraphics[width=0.49\textwidth]{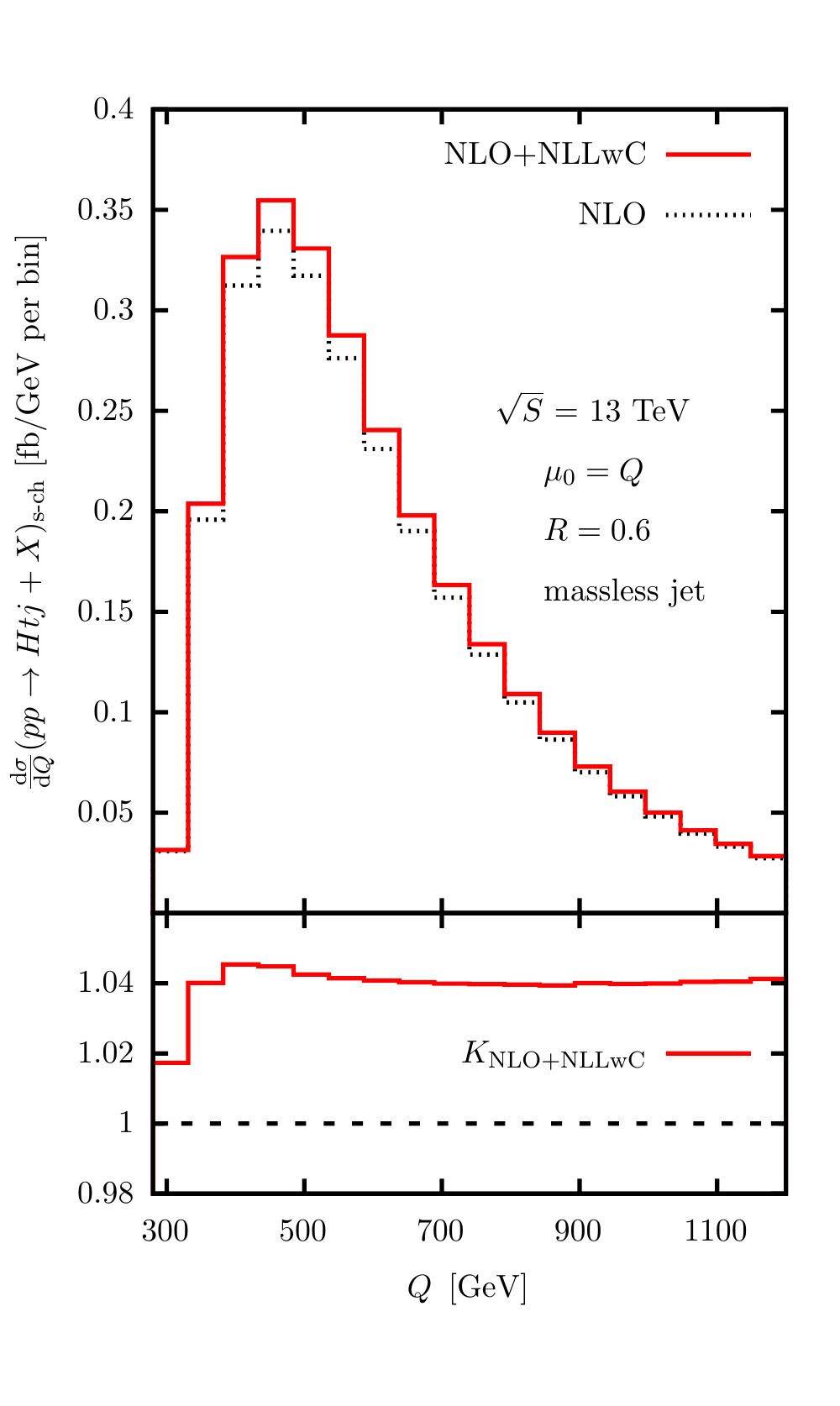}
\caption{Differential distributions in $Q$ for massive (left plot) and massless (right plot) jets for $R=0.6$: (top) absolute values of the NLO, NLO+NLLwC distributions and (bottom) ratio of the NLO+NLLwC result to the NLO prediction.}
\label{fig4}
\end{figure}

Next, we discuss the resummed NLLwC results matched to the NLO predictions. Fig.~\ref{fig4} shows differential distributions in $Q$ for massive and massless jets. We see that the corrections beyond NLO taken into account in the resummed formula amount to 1-4\%, depending on the value of $Q$ and the treatment of the jet. Inspecting the ratios of the NLO+NLLwC to NLO distributions, we also observe that the resummed corrections increase with growing $Q$, especially for massive jets. This behaviour is expected, as soft gluon corrections should be more pronounced closer to the threshold, approched with higher $Q$.

Fig.~\ref{fig5} shows the NLO+NLLwC and NLO total cross sections as functions of the scale ratio $\mu/\mu_0$ for $\mu_0=Q$.  We see that resummation provides a reduction of the scale dependence for both massless and massive jets, with a more noticable effect for massless jets. The higher-order NLLwC corrections modify the absolute value of the total cross section at $\mu=\mu_0=Q$  by up to 3\% percent. The value of these corrections depends on the treatment of the final state jet since $\tilde {J}_{out}^{(1)}$ functions and the resummed final state jet factors $J_k$ differ between the two approaches. Consequently, even if the ${\cal O}(\als)$  NLLwC$|_{\rm NLO} $ expanded results are very close  to each other numerically, cf. Fig.~\ref{fig2},  the resummed cross sections can differ visibly, as they take into account different terms starting from  ${\cal O}(\als^2)$. For example, treating jets as massive leads to resummation of terms with single logarithms of $N$ in the final state jet function. They come together with a $\log \delta$ coefficient, effectively partially resumming also the $\log \delta$ terms. In the massless jet approach, final state jet function does not contain any dependence on $\delta$. 

On the other hand, as demonstrated in Fig.~\ref{fig6},  if the scale variation of results obtained with various central scale choices is considered, the NLO+NLLwC results show much smaller spread in absolute values than the NLO results, even after accounting for the spread due to possible various treatments of the final state jet. It clearly indicates a great potential of resummation methods to increase the stability of the theoretical predictions and motivates further study of the subject.

\begin{figure}
\centering
\includegraphics[width=0.55\textwidth]{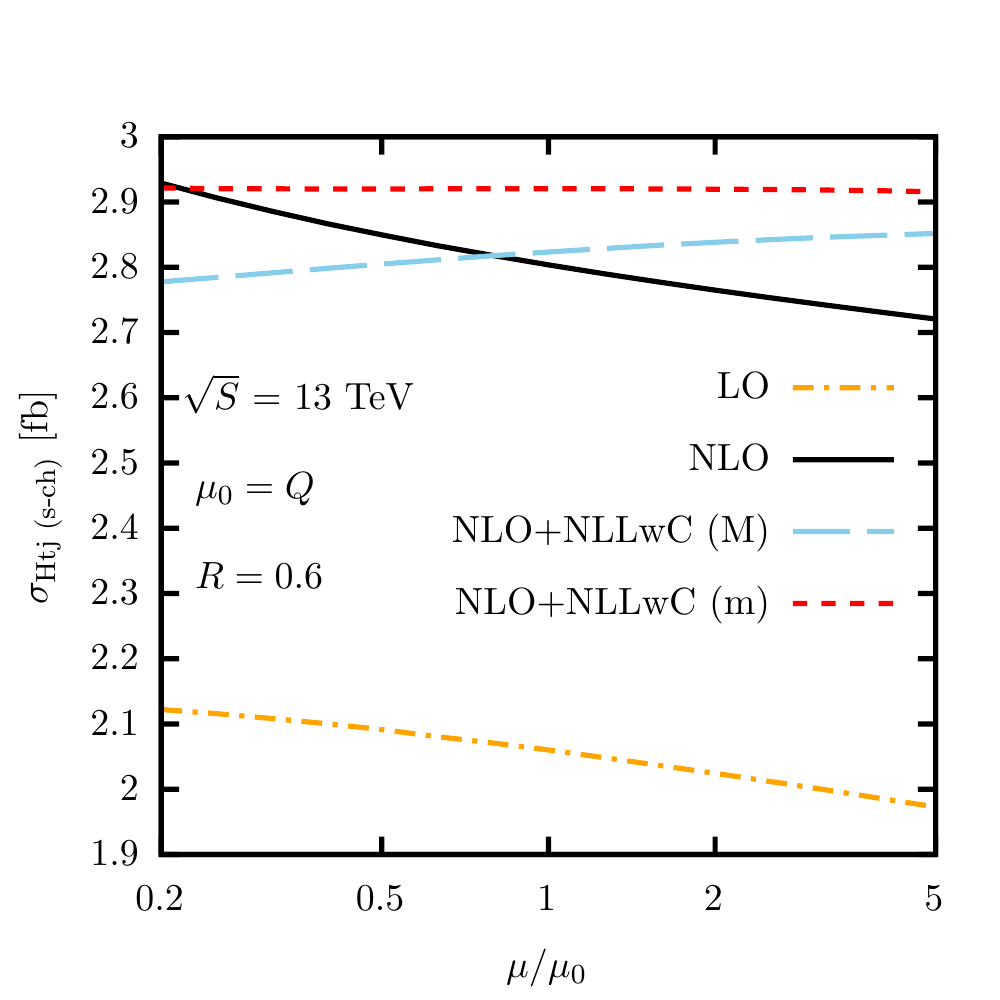}
\caption{Total cross section as a function of $\mu/\mu_0$ for $R=0.6$ at the LO, NLO and the NLO+NLLwC accuracy with the final state jet treated as massive (M) or massless (m).}
\label{fig5}
\end{figure}

\begin{figure}
\centering
\includegraphics[width=0.49\textwidth]{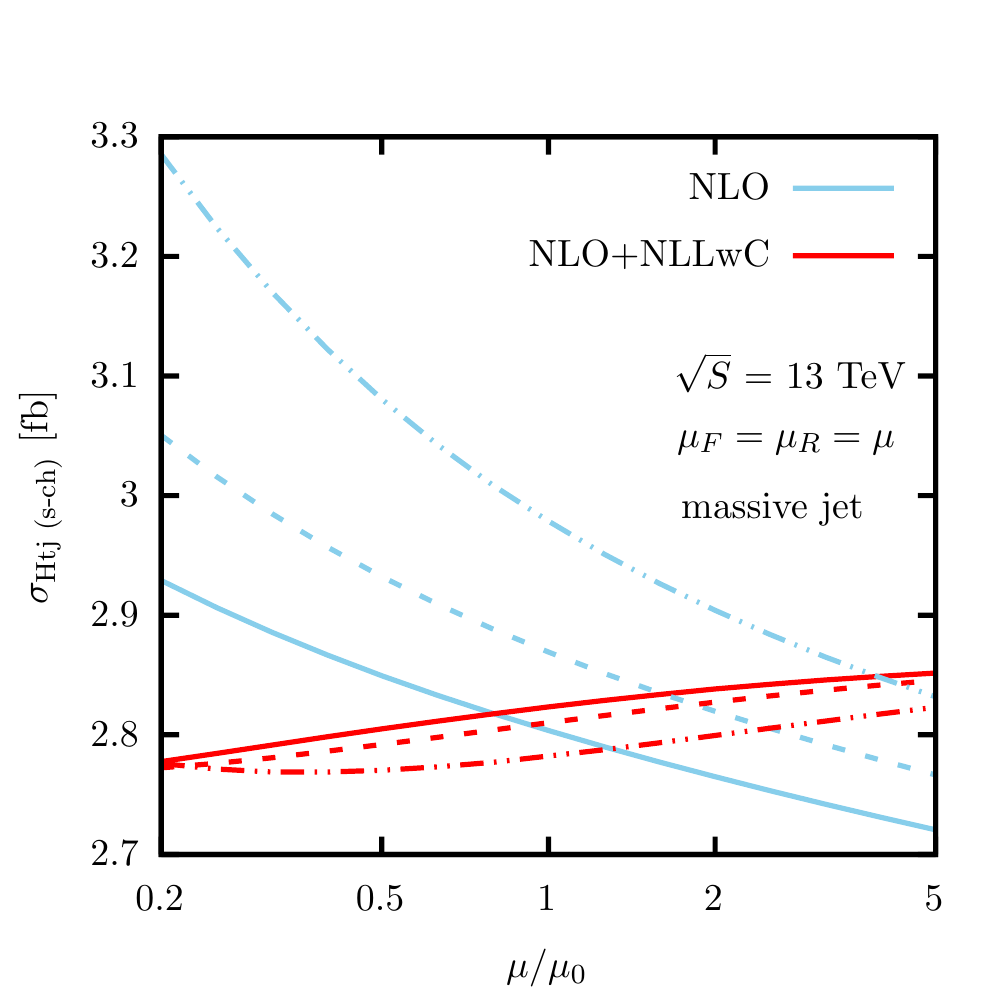}
\includegraphics[width=0.49\textwidth]{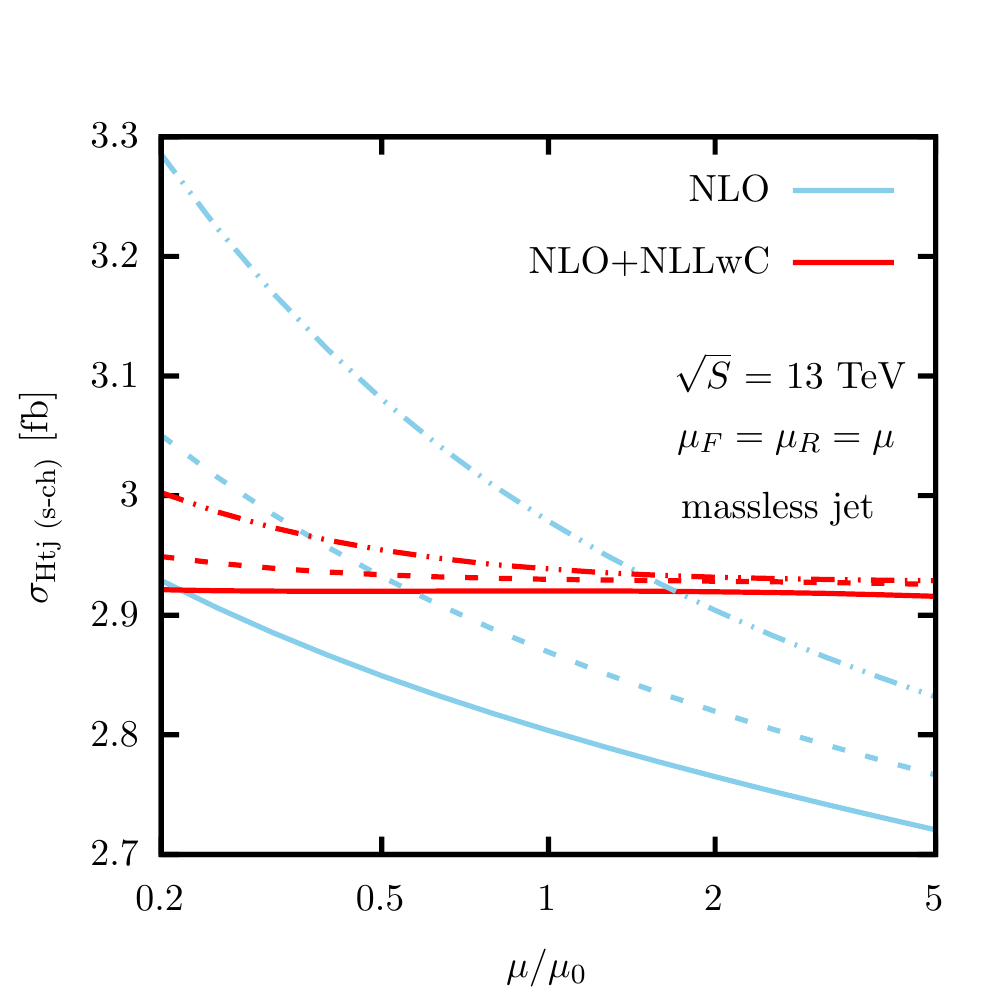}
\caption{Total cross section as a function of $\mu/\mu_0$ for $R=0.6$ at the NLO and NLO+NLLwC accuracy for massive (left plot) and massless (right plot) jets. Results are shown for three central scale choices $\mu_0=Q$ (solid lines), $\mu_0=H_T/2$ (dashed lines), and $\mu_0=H_T/6$ (dashed-dotted lines). }
\label{fig6}
\end{figure}

In our studies, we use a moderate value of the jet radius, $R=0.6$. It can be then expected that logarithmic terms in $R$, or equivalently $\delta$, should not provide large contributions to the cross section. 
While the dependence of the matched resummed results on the assumption regarding the jet mass is not very big, and amounts to 3\% of the total NLO cross section at $\mu_0=Q$, cf. Fig.~\ref{fig5}, it is only slightly smaller than the overall scale uncertainty estimated for $\mu_R=\mu_F$. As we have noted above, the difference between the two results is driven by the $\delta$-dependent terms. Their importance is  heightened by the relative smallness of the pure soft gluon corrections arising in the invariant mass threshold limit for the $tH$ production.  This is particularly visible in the massless jet case, where, as we have discussed, in order to numerically approach the NLO result,  the corrections of ${\cal O}(1/N)$, originating from the region $\delta^2 \bar N \sim 1$, cannot be neglected. While we have included these terms in our calculations, in general they can only generate a part of power-subleading terms at higher order, what may be seen as a drawback of the massless jet approach. In contrast, the NLLwC formula we use for a massive final-state jet does not involve terms formally subleading in $N$. However, this approach does not treat the logarithmic terms in $\delta$ on the same footing, resumming them only partially. At the heart of these intricacies and differences between the two formulations lies a discord between the inclusive treatment of the jet, embodied in the corresponding factorization formula and the notion of the invariant mass threshold, which as we have seen involves some specification of the jet momentum.  While for processes with bigger contributions from soft gluon corrections, i.e. those involving gluons in the initial state, the approach presented here might lead to a very small dependence on the treatment of the jet, our studies show that the case of the $tH$ process requires much more care.

Apart from considering a different threshold variable in a more inclusive kinematics, one could also consider a more exclusive process, i.e. $t$, $H$ and jet in the final state. This would imply a different factorization formula, including non-global logs and separating soft-collinear modes from the global soft and collinear modes, which offers a possibility to sum logarithms of $R$, as demonstrated within the soft-collinear field theory framework~\cite{Liu:2017pbb, Becher:2016mmh, Larkoski:2015zka, Larkoski:2016zzc}. Such an analysis is, however,  beyond the scope of the present paper.

\section{Summary}

The NLO corrections to the $s$-channel $tH$ production are significant, at the level of 35\%. The main contribution to them comes from the quark-initiated production channel, which indicates the importance of higher-order corrections in this channel. The work presented here focuses on exploring ways in which resummation methods in direct QCD resummation can be used to provide an estimate of the higher-order corrections. Of course, as the $tH$ production takes place only in the quark-antiquark channel at LO, the resummation corrections cannot to be expected to be as important as for  the process with gluons in the initial state. However, it is a formally well defined set of higher-order corrections that can be systematically taken into account to all orders and studied numerically. The relevance of the measurement of the $tH$ process for gaining information on the top Yukawa coupling, as well as the increasing experimentall precision make such studies more and more important with time.

In this exploratory work, we investigated a method of threshold resummation in direct QCD, set in the invariant mass kinematics.  The definition of the threshold variable through the invariant mass of the final state system inevitably involves jet's momentum, leading to a residual dependence on a jet-size parameter $R$.  Our numerical results carry typical characteristic of the resummed predictions such as reduced scale dependence in comparison with the fixed-order results. However, it turns out that the pure soft gluon corrections at the invariant mass threshold, i.e. terms logarithmic in $N$, play a relatively small role for the process at hand. This is reflected in the fact that in the two cases we consider here, i.e. treating the final state jet as either massless or massive, quantitatively very similar $\cal{O}(\als)$ corrections  are obtained at different level of accuracy in powers of $N$. Naturally, in such a situation non-logarithmic terms, including also those  dependent on $R$, carry more relevance. In consequence, the resummed results in the two considered cases differ numerically from each other, as they take into account different subsets of higher-order terms.    
To the best of our knowledge, though substantial numerical differences resulting from the treatment of the final state jet have been observed before, e.g. for the single-inclusive jet production~\cite{Kumar:2013hia, deFlorian:2013qia}, this is the first time where this difference is attributed, at least in part, to the terms subleading in powers of $N$. Consequently, our studies indicate the importance of the power-subleading terms in $N$,
underlining a general need for more developments aimed at their systematic all-order treatment. Similarly, our results also indicate that the calculations of the full higher-order corrections are very much needed in order to improve the description of the $tH$ production process.\\

\pagebreak
\noindent
{\Large \bf Acknowledgements}\\
This work has been supported in part by the Deutsche Forschungsgemeinschaft (DFG) grant KU3103/2. V.T. acknowledges funding from the European Union's Horizon 2020 research and innovation programme as part of the Marie Sk\l odowska-Curie Innovative Training Network MCnetITN3 (grant agreement no. 722104), while L.M.V. acknowledges support from the DFG Research Training Group ``GRK 2149: Strong and Weak Interactions - from Hadrons to Dark Matter''.

\bibliographystyle{JHEP}
\providecommand{\href}[2]{#2}\begingroup\raggedright\endgroup

\end{document}